\documentclass{ws-procs9x6}
\usepackage{epsfig}

\def\bm{\boldmath}
\def\mb{\mbox}

\def\bge{\begin{equation}}
\def\ene{\end{equation}}
\def\bg{\begin{eqnarray}}
\def\en{\end{eqnarray}}

\def\ubar{\bar{u}}
\def\dbar{\bar{d}}
\def\sbar{\bar{s}}

\def\D0bar{\overline{D^0}}

\def\vr{\vec{r}}

\begin{document}

\title{\vspace*{-1.5cm} Hadron properties in nuclear medium and their
impacts on observables}

\author{K. Tsushima}

\address{Department of Physics and Astronomy, University of Georgia\\ 
Athens, GA 30602, USA\\ 
E-mail: tsushima@physast.uga.edu}


\maketitle

\abstracts{
The effect of changes in hadron properties in a nuclear 
medium on physical observables is discussed.
Highlighted results are,  
(1) hypernuclei, (2) meosn-nuclear bound states, (3) $K$-meson production 
in heavy ion collisions, and (4) $J/\Psi$ dissociation in a nuclear medium.
In addition, results for the near-threshold 
$\omega$- and $\phi$-meson production in proton proton
collisions are reported.
}

%%%%%%%%%%%%%%%%%%%%%%%%%%%%%%%%%%%%%%%%%%%%%%%%%%%%%%%%%%%%%%%
\section{Hadrons in nuclear medium: 
treatment in QMC~\protect\cite{Guichon,Guichonf,Saito_finite,Tsushima_hyp,Tsushimak,Tsushimaeta,Tsushimad,Tsushimaj} }
%%%%%%%%%%%%%%%%%%%%%%%%%%%%%%%%%%%%%%%%%%%%%%%%%%%%%%%%%%%%%%%

\begin{figure}[th]
\begin{center}
\vspace{-0.4cm}
\psfig{file=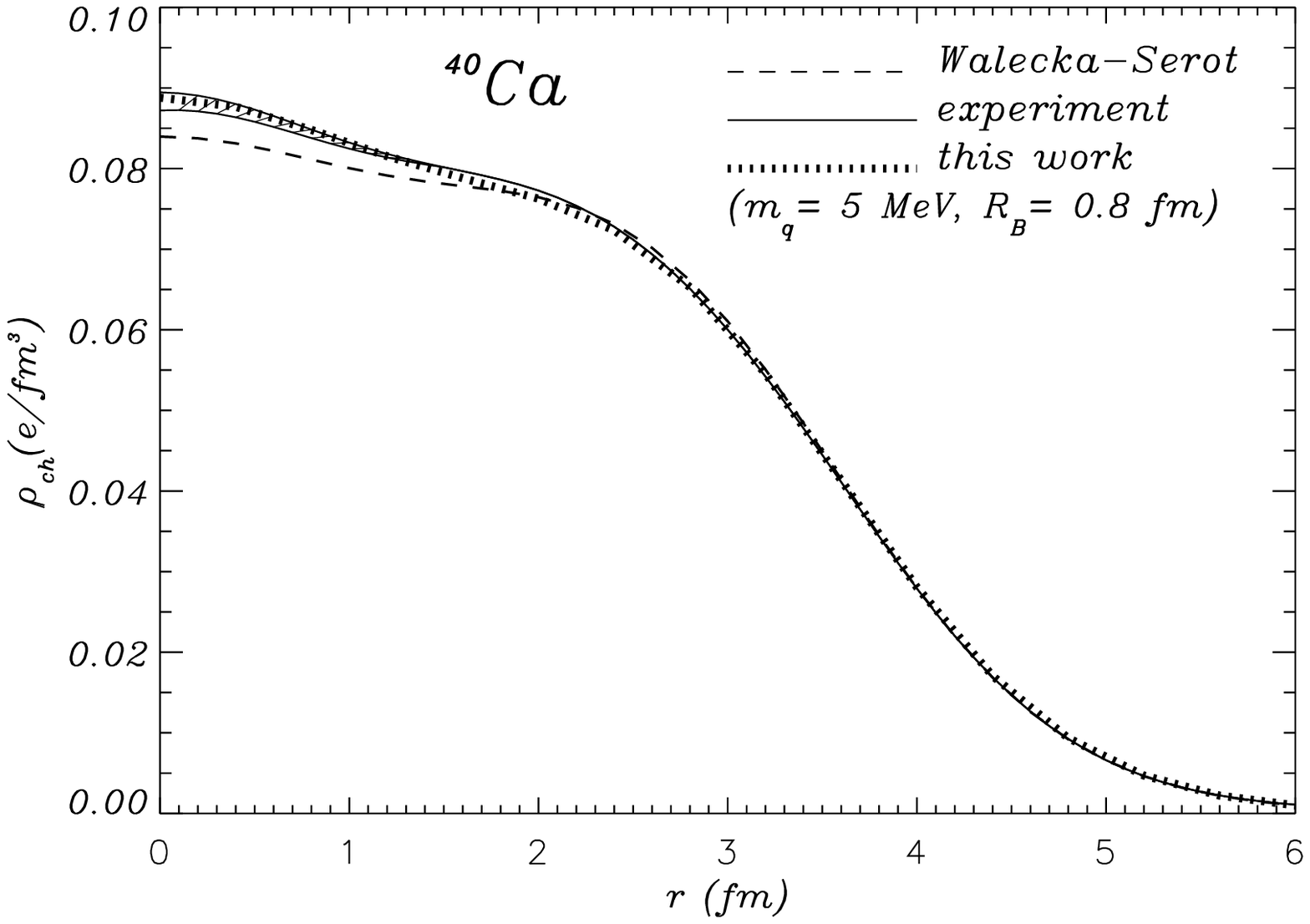,height=4.5cm,width=5.5cm}
\hspace{-0.5cm}
\psfig{file=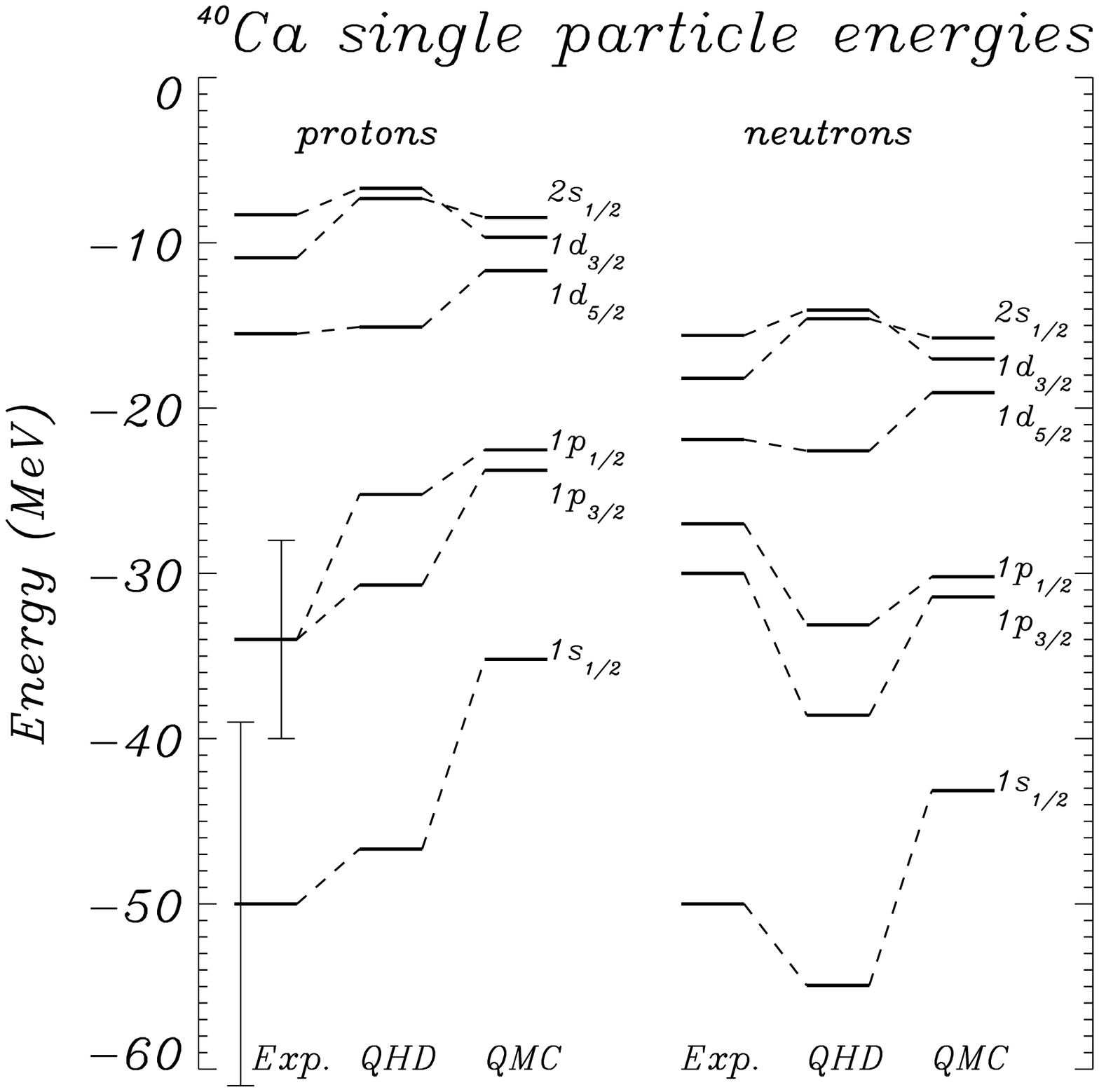,height=4.5cm,width=6.5cm}
\vspace{-0.2cm}
\caption{Charge density distribution 
(the left panel) and energy spectrum 
(the right panel) for $^{40}$Ca nucleus, compared with the experimental 
data and those of Quantum Hadrodynamics 
(QHD, also labeled by \lq\lq Walecka-Serot")~\protect\cite{QHD}.
\label{Ca}
}
\end{center}
\end{figure}

We discuss here hadrons in a nuclear medium in 
quark-meson coupling (QMC) model~\cite{Guichon}.
The model has been extended and successfully applied to many  
problems~\cite{Guichonf,Saito_finite,Tsushima_hyp,Tsushimak,Tsushimaeta,Tsushimad,Tsushimaj}.
A detailed description of the Lagrangian density, the
mean-field equations of motion, and the treatment of finite nuclei 
are given in Refs.~\cite{Guichonf,Saito_finite}.
As examples, we show in Fig.~\ref{Ca} results for $^{40}$Ca nucleus
in QMC~\cite{Saito_finite}, which bases on 
the quark structure of nucleon, or, nucleus.
%
%%%%%%%%%%%%%%%%%%%%%%%%%%%%%%%%%%%%%%%%%%%%%%%%%%%%
\vspace{-0.7cm}
%%%%%%%%%%%%%%%%%%%%%%%%%%%%%%%%%%%%%%%%%%%%%%%%%%%%
%

The Dirac equations for the quarks and antiquarks in hadron bags 
($q = u,\bar{u},d$ or $\bar{d}$, hereafter) 
neglecting the Coulomb force,  
are given by
($|\mbox{\boldmath $x$}|\le$ bag 
radius)~\cite{Guichonf,Saito_finite,Tsushima_hyp,Tsushimak,Tsushimaeta,Tsushimad,Tsushimaj}:
\begin{eqnarray}
\left[ i \gamma \cdot \partial_x -
(m_q - V^q_\sigma)
\mp \gamma^0
\left( V^q_\omega +
\frac{1}{2} V^q_\rho
\right) \right] 
\left( \begin{array}{c} \psi_u(x)  \\
\psi_{\bar{u}}(x) \\ \end{array} \right) &=& 0,
\label{diracu}\\
\left[ i \gamma \cdot \partial_x -
(m_q - V^q_\sigma)
\mp \gamma^0
\left( V^q_\omega -
\frac{1}{2} V^q_\rho
\right) \right]
\left( \begin{array}{c} \psi_d(x)  \\
\psi_{\bar{d}}(x) \\ \end{array} \right) &=& 0,
\label{diracd}\\
%
%\left( (V^q_\omega -\frac{1}{2}V^q_\rho)\, {\rm for}\,
%\left(\begin{array}{c} \psi_d  \\ 
%\psi_{\bar{d}}\\ \end{array} \right) \right),
%\label{diracu}
%
\left[ i \gamma \cdot \partial_x - m_{s,c} \right]
\psi_{s,c} (x)\,\, ({\rm or}\,\, \psi_{\bar{s},\bar{c}}(x)) &=& 0. 
%\psi_{s,c} (x)\,\, ({\rm or}\,\, \psi_{\bar{s},\bar{c}}(x)) = 0, 
%\hspace{8em}
%\left( \to -\frac{1}{2}V^q_\rho\,\, {\rm for}\,\,
%\left(\begin{array}{c} \psi_d  \\ 
%\psi_{\bar{d}}\\ \end{array} \right) \right),
\label{diracsc}
\end{eqnarray}
The mean-field potentials for a bag in nuclear matter
are defined by $V^q_\sigma{\equiv}g^q_\sigma
\sigma$,
$V^q_\omega{\equiv}$ $g^q_\omega
\omega$ and
$V^q_\rho{\equiv}g^q_\rho b$,
with $g^q_\sigma$, $g^q_\omega$ and
$g^q_\rho$ the corresponding quark-meson coupling
constants. 
%%%%
%The normalized, static solution for the ground state quarks or antiquarks
%with flavor $f$ in the hadron, $h$, may be written,  
%
%$\psi_f (x) = N_f e^{- i \epsilon_f t / R_h^*}
%\psi_f (\mbox{\boldmath $x$})$,
%
%where $N_f$ and $\psi_f(\mbox{\boldmath $x$})$
%`are the normalization factor and
%corresponding spin and spatial part of the wave function. 
%%%
The bag radius in medium for a hadron $h$, $R_h^*$, 
will be determined through the
stability condition for the mass of the hadron against the
variation of the bag 
radius~\cite{Guichon,Guichonf}
(see Eq.~(\ref{hmass})). 
%%%%%
%The eigenenergies  
%in units of $1/R_h^*$ are given by
%
%\bge
%\left( \begin{array}{c}
%\epsilon_u \\
%\epsilon_{\bar{u}}
%\end{array} \right)
%= \Omega_q^* \pm R_h^* \left(
%V^q_\omega
%+ \frac{1}{2} V^q_\rho \right),\,\,
%(V^q_\omega - \frac{1}{2} V^q_\rho {\rm\,\, for\,\,} d, \dbar),\,\,
%
%\left( \begin{array}{c} \epsilon_d \\
%\epsilon_{\bar{d}}
%\end{array} \right)
%= \Omega_q^* \pm R_h^* \left(
%V^q_\omega
%- \frac{1}{2} V^q_\rho \right),
%\non\\
%
%\epsilon_{s,c}
%= \epsilon_{\bar{s},\bar{c}} =
%\Omega_{s,c},
%\label{cenergy}
%\ene
%
%%%%
The hadron masses
in a nuclear medium $m^*_h$ (free mass will be denoted by $m_h$),
are calculated by
\begin{eqnarray}
%
%%%%%%
%m_h^* &=& \frac{ n_q\Omega_q^*
%+ n_{s,c}\Omega_{s,c} - z_h}{R_h^*}
%+ {4\over 3}\pi R_h^{* 3} B,\quad
%%%%%%%
m_h^* &=& \sum_{j=q,\bar{q},Q,\overline{Q}} 
\frac{ n_j\Omega_j^* - z_h}{R_h^*}
+ {4\over 3}\pi R_h^{* 3} B,\quad
\left. \frac{\partial m_h^*}
{\partial R_h}\right|_{R_h = R_h^*} = 0,
\label{hmass}
\end{eqnarray}
where $\Omega_q^*=\Omega_{\bar{q}}^*
=[x_q^2 + (R_h^* m_q^*)^2]^{1/2}\,(q=u,d)$, with
$m_q^*=m_q{-}g^q_\sigma \sigma$,
$\Omega_Q^*=\Omega_{\overline{Q}}^*=[x_Q^2 + (R_h^* m_Q)^2]^{1/2}\,(Q=s,c)$,
and $x_{q,Q}$ being the bag eigenfrequencies.
$B$ is the bag constant, $n_q (n_{\bar{q}})$ and $n_Q (n_{\overline{Q}})$ 
are the lowest mode quark (antiquark) 
numbers for the quark flavors $q$ and $Q$ 
in the hadron $h$, respectively, 
and the $z_h$ parametrize the sum of the
center-of-mass and gluon fluctuation effects (assumed to be
independent of density). 
%%%%%%%%%%%%%%%%%%%%%%%%%%%%%%%%%%%%%%%%%%%%%%%%%%%%
\begin{figure}[th]
\begin{center}
\vspace{-0.6cm}
\psfig{file=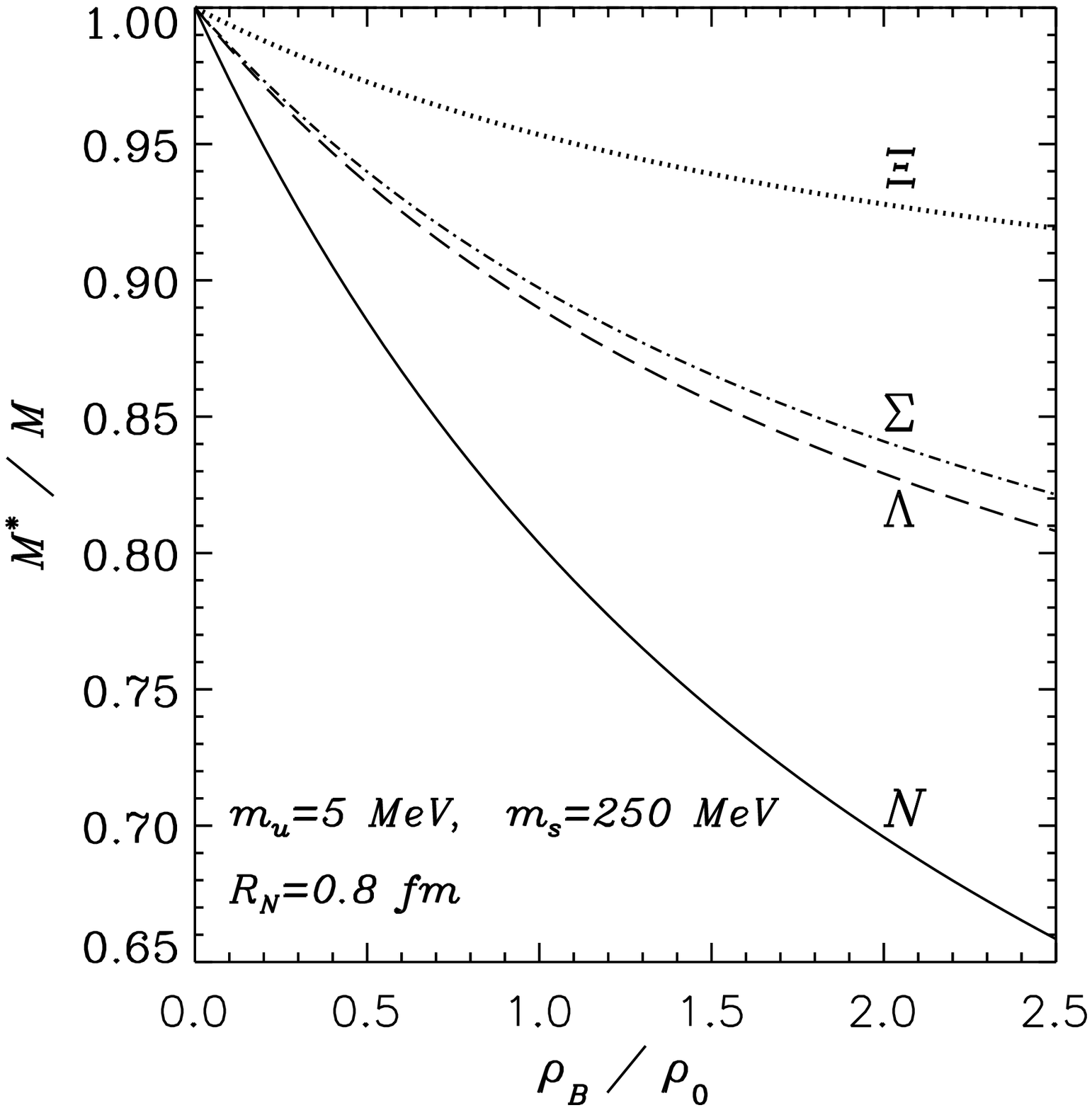,height=4.2cm,width=6cm}
\hspace{-0.5cm}
\psfig{file=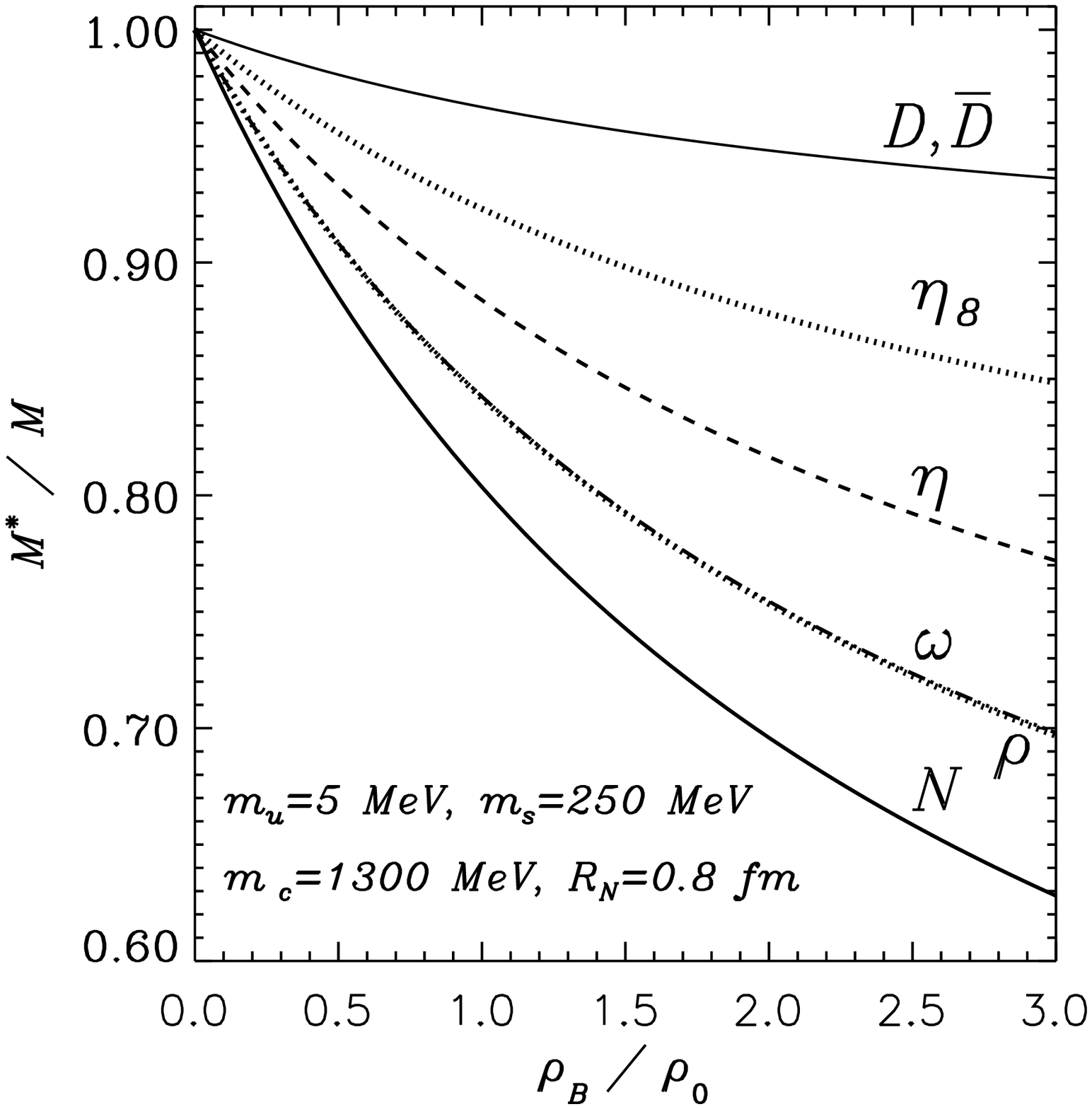,height=4.2cm,width=6cm}
\vspace{-0.3cm}
\psfig{file=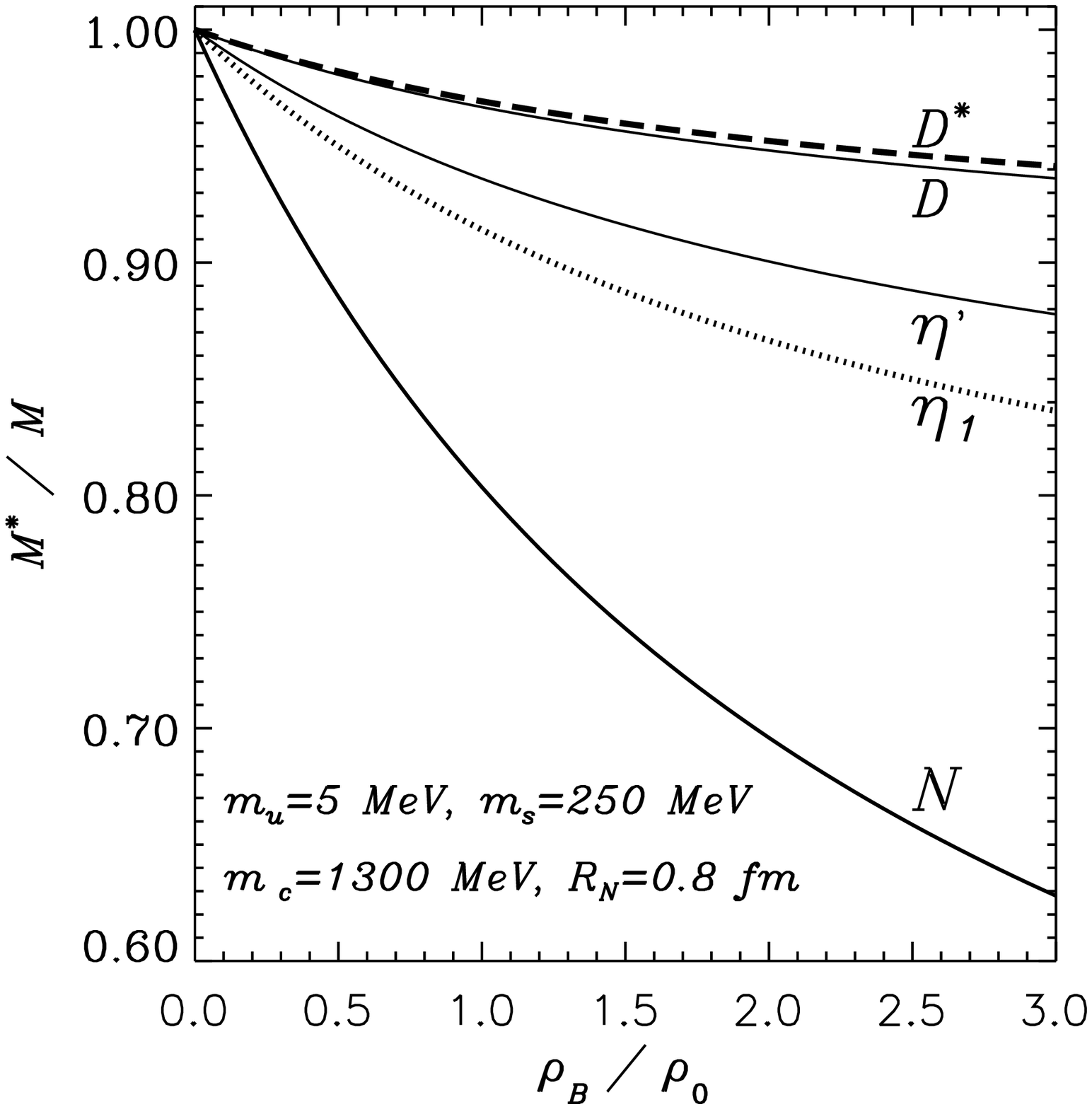,height=4.2cm,width=6cm}
\hspace{-0.5cm}
\psfig{file=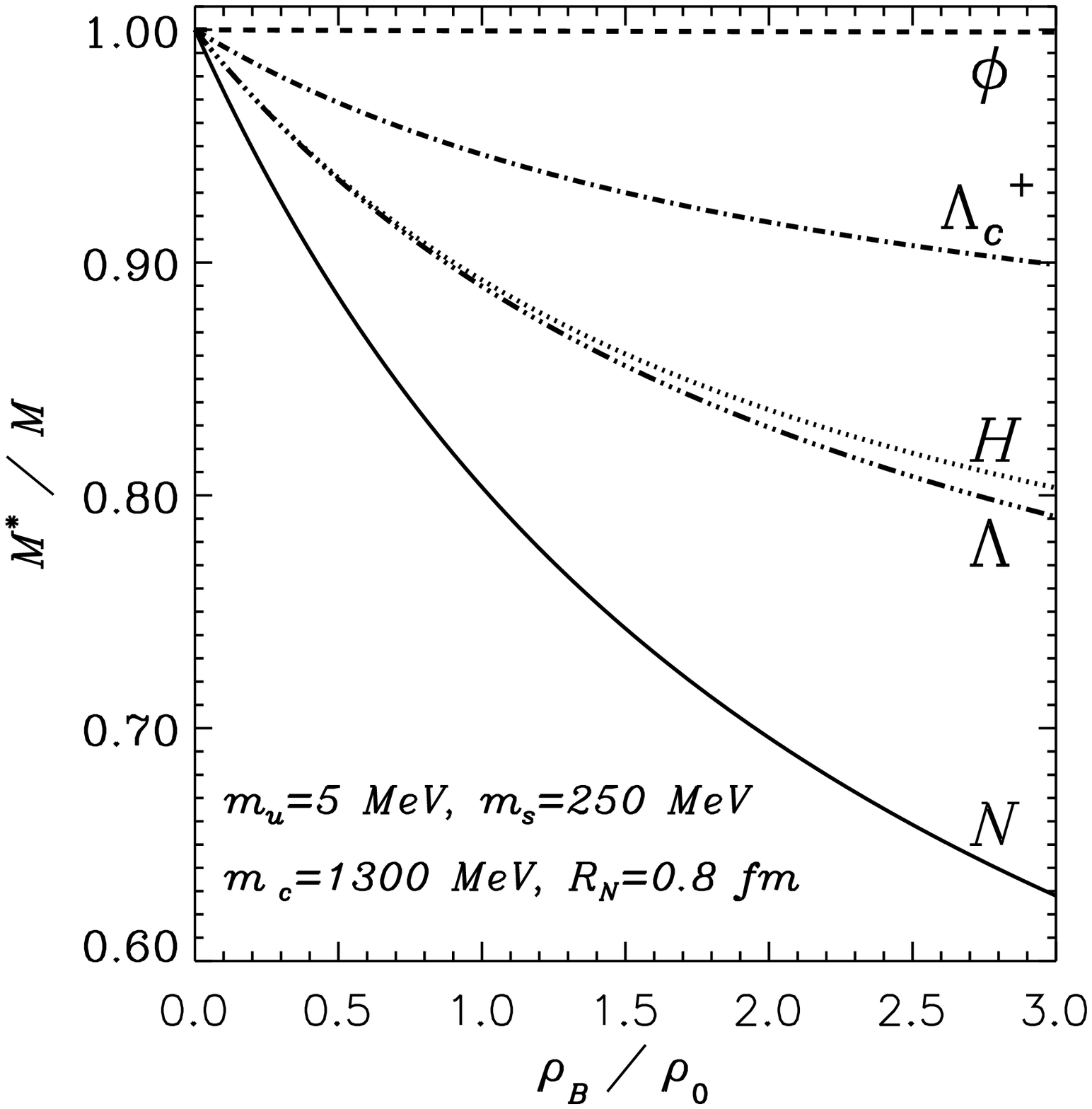,height=4.2cm,width=6cm}
\vspace{-0.3cm}
\psfig{file=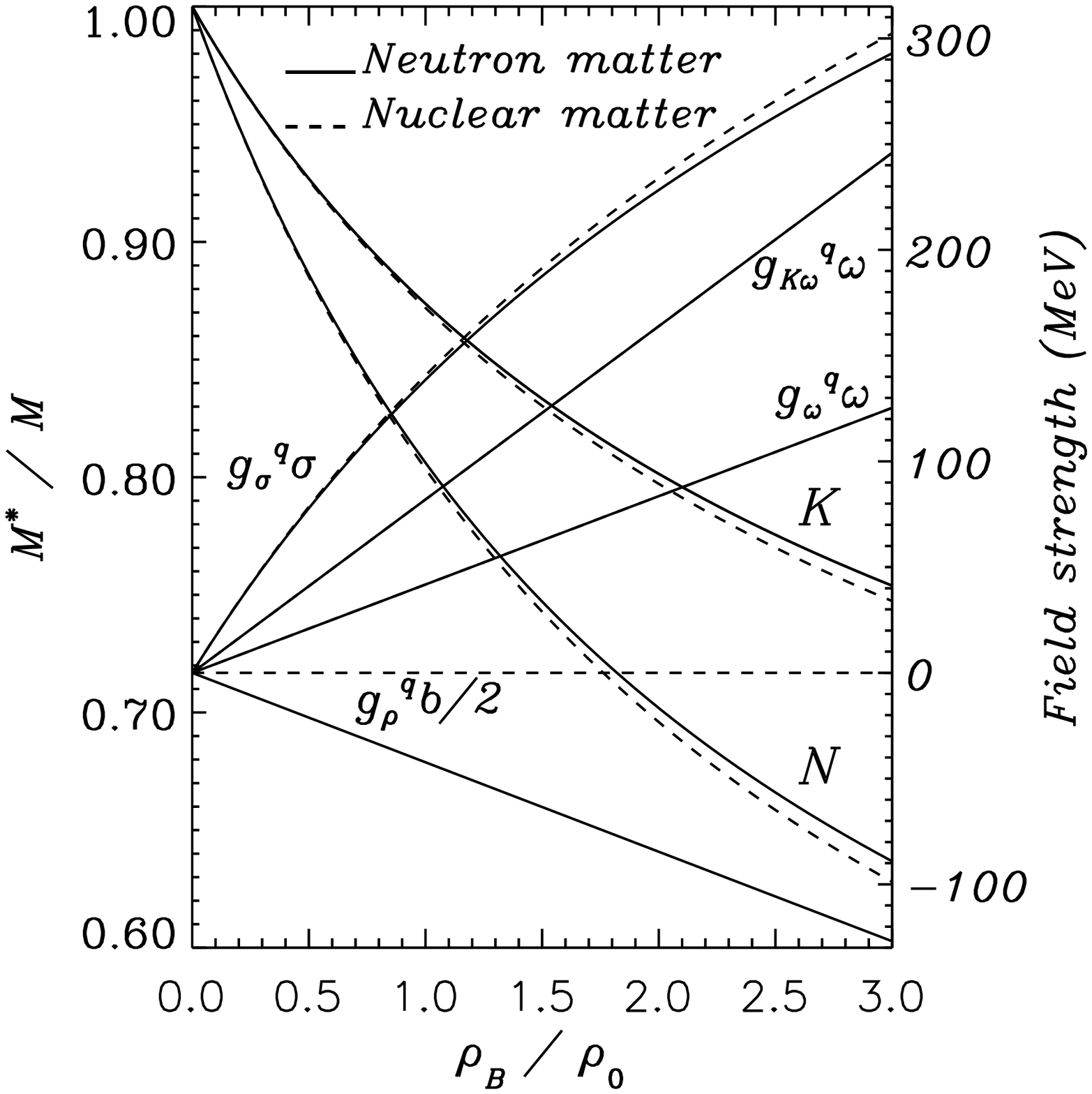,height=4.2cm,width=5.5cm}
\hspace{0.5cm}
\psfig{file=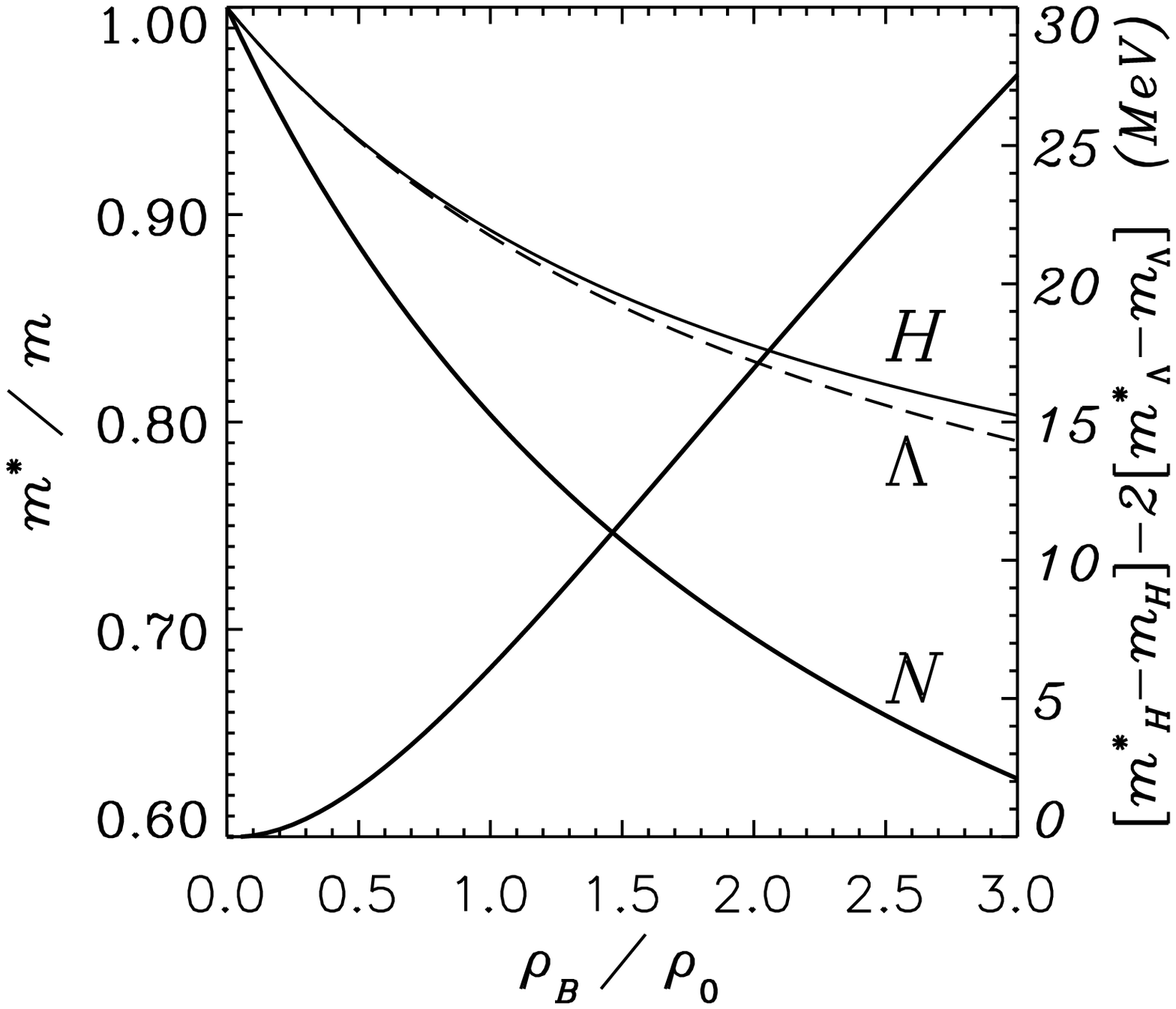,height=4.2cm,width=5.0cm}
%\vspace{-0.2cm}
\caption{Effective mass ratios 
and mean field potentials for hadrons
in nuclear matter ($\rho_0=0.15$ fm$^{-3}$). 
$\lq\lq H"$ stands for the H particle with an input $m_H=2m_\Lambda$.
\label{effective_mass}
\vspace{-0.5cm}
}
\end{center}
\end{figure}
%%%%%%%%%%%%%%%%%%%%%%%%%%%%%%%%%%%%%%%%%%%%%%%%%%%%
The parameters are determined in free space to
reproduce the corresponding masses.
We chose the values,
$(m_q, m_s, m_c) = (5, 250, 1300)$ MeV for the
current quark masses, and $R_N = 0.8$
fm for the bag radius of the nucleon in free space. 
The quark-meson coupling constants, $g^q_\sigma$, $g^q_\omega$
and $g^q_\rho$, are adjusted to fit the nuclear 
saturation energy and density of symmetric nuclear matter, and the bulk
symmetry energy~\cite{Guichonf}. 
%%%%%
%Exactly the same coupling constants, $g^q_\sigma$, $g^q_\omega$ and
%$g^q_\rho$, are used for the light quarks in the mesons and hyperons as in 
%the nucleon. 
%%%%%
However, in studies of the kaon system, we found that it was
phenomenologically necessary to increase the strength of the vector
coupling to the non-strange quarks in the $K^+$ (by a factor of
$1.4^2$, i.e., $g_{K\omega}^q \equiv 1.4^2 g^q_\omega$) 
in order to reproduce the empirically extracted $K^+$-nucleus
interaction~\cite{Tsushimak}.  
We assume this also for the $D$ and $\overline{D}$ 
mesons~\cite{Tsushimad,Tsushimaj,Alex_charm}.
The scalar ($V^{h}_s$) and vector ($V^{h}_v$) potentials 
felt by the hadrons $h$,  
in nuclear matter are given by,
\begin{equation}
V^h_s = m^*_h - m_h,\,\,
V^h_v =
  (n_q - n_{\bar{q}}) {V}^q_\omega - I_3 V^q_\rho 
\,\, (V^q_\omega \to 1.4^2 {V}^q_\omega\,\, 
{\rm for}\, K,\overline{K},D,\overline{D}), 
\label{svpot}
\end{equation}
where $I_3$ is the third component of isospin projection  
of the hadron $h$. 

In Fig.~\ref{effective_mass} we show effective masses and 
mean field potentials 
for various hadrons in symmetric nuclear matter.
Several comments are on the results shown in Fig.~\ref{effective_mass}:
\begin{enumerate}
\item{Physical $\eta$ and $\eta'$ mesons are treated including mixing angle,
$\theta_P= -10^{\circ}$, 
as ($\eta, \eta'$) = ($\eta_8\cos\theta_P - \eta_1\sin\theta_P$,
$\eta_8\sin\theta_P + \eta_1\cos\theta_P$), where
$\eta_1=(1/\sqrt{3})(u\ubar + d\dbar + s\sbar)$ and
$\eta_8=(1/\sqrt{6})(u\ubar + d\dbar - 2s\sbar)$.}
\item{The H particle is treated as a 6-quark bag, $uuddss$, with 
$m_H = 2m_\Lambda$.
}
\item{The scalar potential for the hadron $h$,
$V_s^h$, shows a universal quark number scaling rule,
$V_s^h/V_s^N \simeq (n_q+n_{\bar{q}})/3$, 
where $V_s^N$ is the scalar potential for the nucleon.
(See Eq.~(\ref{svpot}).)}
\item{The scalar potential for the $\phi$ meson arises entirely 
from the $\phi - \omega$ mixing in QMC, and tiny.}
\item{Reduction in effective mass for the $\Lambda_c^+$ 
implies the role of the light quarks in partial restoration of 
chiral symmetry in a heavy-light quark system, 
and has a merit of further study.}
\end{enumerate}
In following sections we will discuss how these changes 
in hadron properties in a nuclear medium
give impacts on physical observables.

%%%%%%%%%%%%%%%%%%%%%%%%%%%%%%%%%%%%%%%%%%%%%%%%%%%%%%%%%%%%%%%%%%%%%%%%%%%
\section{Hypernuclei~\protect\cite{Tsushima_hyp}}
%%%%%%%%%%%%%%%%%%%%%%%%%%%%%%%%%%%%%%%%%%%%%%%%%%%%%%%%%%%%%%%%%%%%%%%%%%%

When a hyperon, $Y$, is embedded in a nucleus,  
the potential felt by the hyperon at a given position of the nucleus 
can be calculated self-consistently~\cite{Tsushima_hyp}.
Simultaneously one can get also  
energy levels for the hyperon $Y$ in a  
shell-model calculation~\cite{Tsushima_hyp}.
Because the treatment is based on the quark degrees of freedom, 
we need to consider a possibility of Pauli blocking effect 
at the quark level, and also effect of $\Sigma$-$\Lambda$ channel coupling 
for the $\Sigma$- and $\Lambda$-hypernuclei. 
We included these effects in a phenomenological way~\cite{Tsushima_hyp}.
Nevertheless, we can study all 
$\Lambda$-, $\Sigma$-, $\Xi$-hypernuclei in a systematic way,  
due to the quark degrees of freedom, because the hyperon-meson 
coupling constants in nuclear medium are automatically 
determined~\cite{Tsushima_hyp}.
In Table~\ref{hyper} we present QMC predictions for the energy levels 
of hyperons in various hypernuclei.
Results for other heavier mass hypernuclei are given in 
Ref.~\cite{Tsushima_hyp}.

%
%%%%%%%%%%%%%%%%%%%%%%%%%%%%%%%%%%%%%%%%%%%%%%%%%%%%%%%%%%%%%%%%%%%%%%%%%%%%%
%^17_YO
\vspace{-0.2cm}
\begin{table}[h]
\begin{center}
\tbl{Energy levels for hyperons $Y$ (in MeV), 
for $^{17}_Y$O, $^{41}_Y$Ca and $^{49}_Y$Ca hypernuclei, calculated in QMC.
Experimental data are taken from Ref.~\protect\cite{hypdata}.}
{\footnotesize
\begin{tabular}[t]{c|ccccccc}
\hline \hline
&$^{16}_\Lambda$O (Expt.)
&$^{17}_\Lambda$O    &$^{17}_{\Sigma^-}$O
&$^{17}_{\Sigma^0}$O &$^{17}_{\Sigma^+}$O
&$^{17}_{\Xi^-}$O    &$^{17}_{\Xi^0}$O\\
\hline \hline
$1s_{1/2}$&-12.5      &-14.1 &-17.2 &-9.6  &-3.3  &-9.9  &-4.5 \\
$1p_{3/2}$&           &-5.1  &-8.7  &-3.2  &---   &-3.4  &---  \\
$1p_{1/2}$&-2.5 ($1p$)&-5.0  &-8.0  &-2.6  &---   &-3.4  &---  \\ 
\hline \hline
&$^{40}_\Lambda$Ca (Expt.)
&$^{41}_\Lambda$Ca    &$^{41}_{\Sigma^-}$Ca
&$^{41}_{\Sigma^0}$Ca &$^{41}_{\Sigma^+}$Ca
&$^{41}_{\Xi^-}$Ca    &$^{41}_{\Xi^0}$Ca\\
\hline \hline
$1s_{1/2}$&-20.0       &-19.5 &-23.5 &-13.4 &-4.1  &-17.0 &-8.1 \\
$1p_{3/2}$&            &-12.3 &-17.1 &-8.3  &---   &-11.2 &-3.3 \\
$1p_{1/2}$&-12.0 ($1p$)&-12.3 &-16.5 &-7.7  &---   &-11.3 &-3.4 \\
$1d_{5/2}$&            &-4.7  &-10.6 &-2.6  &---   &-5.5  &---  \\
$2s_{1/2}$&            &-3.5  &-9.3  &-1.2  &---   &-5.4  &---  \\
$1d_{3/2}$&            &-4.6  &-9.7  &-1.9  &---   &-5.6  &---  \\ 
\hline \hline
&---
&$^{49}_\Lambda$Ca    &$^{49}_{\Sigma^-}$Ca
&$^{49}_{\Sigma^0}$Ca &$^{49}_{\Sigma^+}$Ca
&$^{49}_{\Xi^-}$Ca    &$^{49}_{\Xi^0}$Ca\\
\hline \hline
$1s_{1/2}$&           &-21.0 &-19.3 &-14.6 &-11.5 &-14.7 &-12.0\\
$1p_{3/2}$&           &-13.9 &-11.4 &-9.4  &-7.5  &-8.7  &-7.4 \\
$1p_{1/2}$&           &-13.8 &-10.9 &-8.9  &-7.0  &-8.8  &-7.4 \\
$1d_{5/2}$&           &-6.5  &-5.8  &-3.8  &-2.0  &-3.8  &-2.1 \\
$2s_{1/2}$&           &-5.4  &-6.7  &-2.6  &---   &-4.6  &-1.1 \\
$1d_{3/2}$&           &-6.4  &-5.2  &-3.1  &-1.2  &-3.8  &-2.2 \\
$1f_{7/2}$&           &---   &-1.2  &---   &---   &---   &---  \\
\hline \hline
\end{tabular}\label{hyper}}
\end{center}
%%%kkkkkkkkk
%\vspace{-0.5cm}
\vspace{-0.7cm}
\end{table}
%
%%%%%%%%%%%%%%%%%%%%%%%%%%%%%%%%%%%%%%%%%%%%%%%%%%%%%%%%%%%%%%%%%%%%%%%
%

%%%%%%%%%%%%%%%%%%%%%%%%%%%%%%%%%%%%%%%%%%%%%%%%%%%%%%%%%%%%%%%%%%%%%%%%%%%
\section{Meson-nuclear bound states~\protect\cite{Tsushimaeta,Tsushimad}} 
%%%%%%%%%%%%%%%%%%%%%%%%%%%%%%%%%%%%%%%%%%%%%%%%%%%%%%%%%%%%%%%%%%%%%%%%%%%

Next, we discuss the meson-nuclear bound states.
We have solved the Klein-Gordon equation for mesons $j$ 
($j=\omega,\eta,\eta',D,\overline{D}$) with the situation of almost 
zero momenta, 
using the calculated potentials in QMC~\cite{Tsushimaeta,Tsushimad}:
\bg
&&\left[ \nabla^2 + E^{*2}_j - \tilde{m}^{*2}_j(r) \right]\,
\phi_j(\vr) = 0,\quad E^*_j \equiv E_j + m_j - i \Gamma_j/2,  
\\
&&\tilde{m}^*_j(r) \equiv
m^*_j(r) - \frac{i}{2}
\left[ (m_j - m^*_j(r))
\gamma_j + \Gamma_j \right]
\equiv m^*_j(r) - \frac{i}{2} \Gamma^*_j (r),
\label{width}
\en
where $E^*_j$ is the complex valued, total energy  
of the meson, and we included
the widths of the mesons in a nucleus assuming a specific form
using $\gamma_j$, which are treated as phenomenological
parameters.
%%%%%
%According to the estimates in Refs.~\cite{hayano,Friman},
%the widths of the mesons in nuclei and at normal nuclear matter density
%are $\Gamma^*_\eta \sim 30 - 70$ MeV~\cite{hayano}
%and $\Gamma^*_\omega \sim 30 - 40$ MeV~\cite{Friman}, respectively.
%%%%
We calculate the single-particle energies for the values
$\gamma_\omega = 0.2$, and $\gamma_\eta = 0.5$, which are
expected to correspond best with experiments~\cite{Tsushimaeta}, while
for the $\eta'$, $D$ and $\overline{D}$, the widths $\Gamma^*_j = 0$ 
are assumed~\cite{Tsushimad}.
For a comparison we present also results for the $\omega$ calculated 
using the potentials obtained in QHD~\cite{Saitoomega}.
Results are given in Tables~\ref{mesic} and~\ref{D_mesic}.
%
%%%%%%%%%%%%%%%%%%%%%%%%%%%%%%%%%%%%%%%%%%%%%%%%%%%
%Omega
\begin{table}[ht]
\vspace{-0.2cm}
\begin{center}
\tbl{Calculated $\omega$-, $\eta$- and $\eta'$-nuclear bound
state energies (in MeV),
$E_j = Re (E^*_j - m_j)\,(j=\omega,\eta,\eta')$,
in QMC~\protect\cite{Tsushimaeta} and those for the $\omega$ in
QHD with $\sigma$-$\omega$ mixing effect~\protect\cite{Saitoomega}.
The complex eigenenergies are given by,
$E^*_j = E_j + m_j - i \Gamma_j/2$.
\qquad (* not calculated)}
{\footnotesize
\begin{tabular}[t]{lc|cc||c||cc|cc}
\hline \hline
& &$\gamma_\eta=0.5$ &(QMC) &(QMC) &$\gamma_\omega=0.2$
&(QMC) &$\gamma_\omega=0.2$ &(QHD)\\
\hline \hline
& &$E_\eta$ &$\Gamma_\eta$ &$E_{\eta'}$ &$E_\omega$ &$\Gamma_\omega$
&$E_\omega$ &$\Gamma_\omega$\\
\hline
$^{6}_j$He &1s &-10.7&14.5 & * &-55.6&24.7 &-97.4&33.5 \\
\hline
$^{11}_j$B &1s &-24.5&22.8 & * &-80.8&28.8 &-129&38.5 \\
\hline
$^{26}_j$Mg &1s &-38.8&28.5 & * &-99.7&31.1 &-144&39.8 \\
            &1p &-17.8&23.1 & * &-78.5&29.4 &-121&37.8 \\
            &2s & --- & --- & * &-42.8&24.8 &-80.7&33.2  \\
\hline \hline
$^{16}_j$O &1s &-32.6&26.7 &-41.3 &-93.4&30.6 &-134&38.7 \\
           &1p &-7.72&18.3 &-22.8 &-64.7&27.8 &-103&35.5 \\
\hline
$^{40}_j$Ca &1s &-46.0&31.7 &-51.8 &-111&33.1  &-148&40.1 \\
            &1p &-26.8&26.8 &-38.5 &-90.8&31.0 &-129&38.3 \\
            &2s &-4.61&17.7 &-21.9 &-65.5&28.9 &-99.8&35.6  \\
\hline
$^{90}_j$Zr &1s &-52.9&33.2 &-56.0 &-117&33.4  &-154&40.6 \\
            &1p &-40.0&30.5 &-47.7 &-105&32.3  &-143&39.8 \\
            &2s &-21.7&26.1 &-35.4 &-86.4&30.7 &-123&38.0 \\
\hline
$^{208}_j$Pb &1s &-56.3&33.2 &-57.5 &-118&33.1 &-157&40.8 \\
             &1p &-48.3&31.8 &-52.6 &-111&32.5 &-151&40.5 \\
             &2s &-35.9&29.6 &-44.9 &-100&31.7 &-139&39.5 \\
\hline \hline
\end{tabular}\label{mesic}}
\end{center}
\vspace{-0.6cm}
\end{table}
%%%%%%%%%%%%%%%%%%%%%%%%%%%%%%%%%%%%%%%%%%%%%%%%%%%
%
%%%%%%%%%%%%%%%%%%%%%%%%%%%%%%%%%%%%%%%%%%%%%%%%%%%%%%%%%%%%%%%%%%%%%%%%%%%%%
\begin{table}[ht]
\vspace*{-0.2cm}
\begin{center}
\tbl{$D^-$, $\D0bar$ and $D^0$ bound state energies (in MeV).
The widths are all set to zero.}
{\footnotesize 
\begin{tabular}[t]{lcccccc}
\hline\hline
state  &$D^- (1.4^2V^q_\omega)$ &$D^- (V^q_\omega)$
&$D^- (V^q_\omega$, no Coulomb) &$\D0bar (1.4^2V^q_\omega)$
&$\D0bar (V^q_\omega)$ &$D^0 (V^q_\omega)$ \\
\hline\hline
%$^{208}_{\bar{D},D^0}$Pb &
                         1s &-10.6 &-35.2 &-11.2 &unbound &-25.4 &-96.2\\
                         1p &-10.2 &-32.1 &-10.0 &unbound &-23.1 &-93.0\\
                         2s & -7.7 &-30.0 & -6.6 &unbound &-19.7 &-88.5\\
\hline\hline
\end{tabular}\label{D_mesic}}
\end{center}
\vspace{-0.4cm}
\end{table}
%%%%%%%%%%%%%%%%%%%%%%%%%%%%%%%%%%%%%%%%%%%%%%%%%%%%%%%%%%%%%%%%%%%%%%%%%%%

Our results suggest that $\omega$, $\eta$ and  $\eta'$ mesons should be
bound in all the nuclei considered. Furthermore, the $D^-$ meson
should be bound in $^{208}$Pb in any case, 
assisted by the Coulomb force~\cite{Tsushimad}.
%%%%%%
%The existence of any bound states at all would give us important
%information concerning the role of the Lorentz scalar $\sigma$ field,
%and hence dynamical symmetry breaking.
%%%%%%%%%%

%%%%%%%%%%%%%%%%%%%%%%%%%%%%%%%%%%%%%%%%%%%%%%%%%%%%%%%%%%%%%%%%%%%%%%%%%%%
\section{\mb\bm{$K$}-meson production in heavy ion 
collisions~\protect\cite{kprod0,kprod_medium}}
\begin{figure}[th]
\begin{center}
\vspace{-2.2cm}
\psfig{file=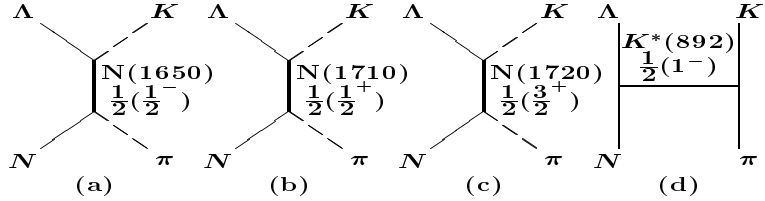,height=11cm,width=11cm}
\vspace{-6.5cm}
\caption{Processes included for the $\pi N \to \Lambda K$ reactions.
\label{k1}
}
\end{center}
\vspace{-0.8cm}
\end{figure}

%%%%%%%%%%%%%%%%%%%%%%%%%%%%%%%%%%%%%%%%%%%%%%%%%%%%%%%%%%%%%%%%%%%%%%%%%%%
%
%Since in  heavy ion collisions at
%SIS energies~\cite{Barth,Schroter}
%the $K^+$-mesons are predominantly produced by secondary pions,
%%%%
Here our focus is the kaon production reactions,
$\pi N \to \Lambda K$, 
in nuclear matter. We calculate the in-medium
reaction amplitudes, taking into account the scalar and vector
potentials for incident, final and intermediate mesons and baryons.
The processes considered in the calculation, which were
already established in studies for 
free space~\cite{kprod0}, are shown in Fig.~\ref{k1}.
%%%%%%%%%%%%%%%%%%%%%%%%%%%%%%%%%%%%%%%%%%%%%%%%%%%%
%%%%%%%%%%%%%%%%%%%%%%%%%%%%%%%%%%%%%%%%%%%%%%%%%%
\begin{figure}[th]
\begin{center}
\vspace{-0.5cm}
\psfig{file=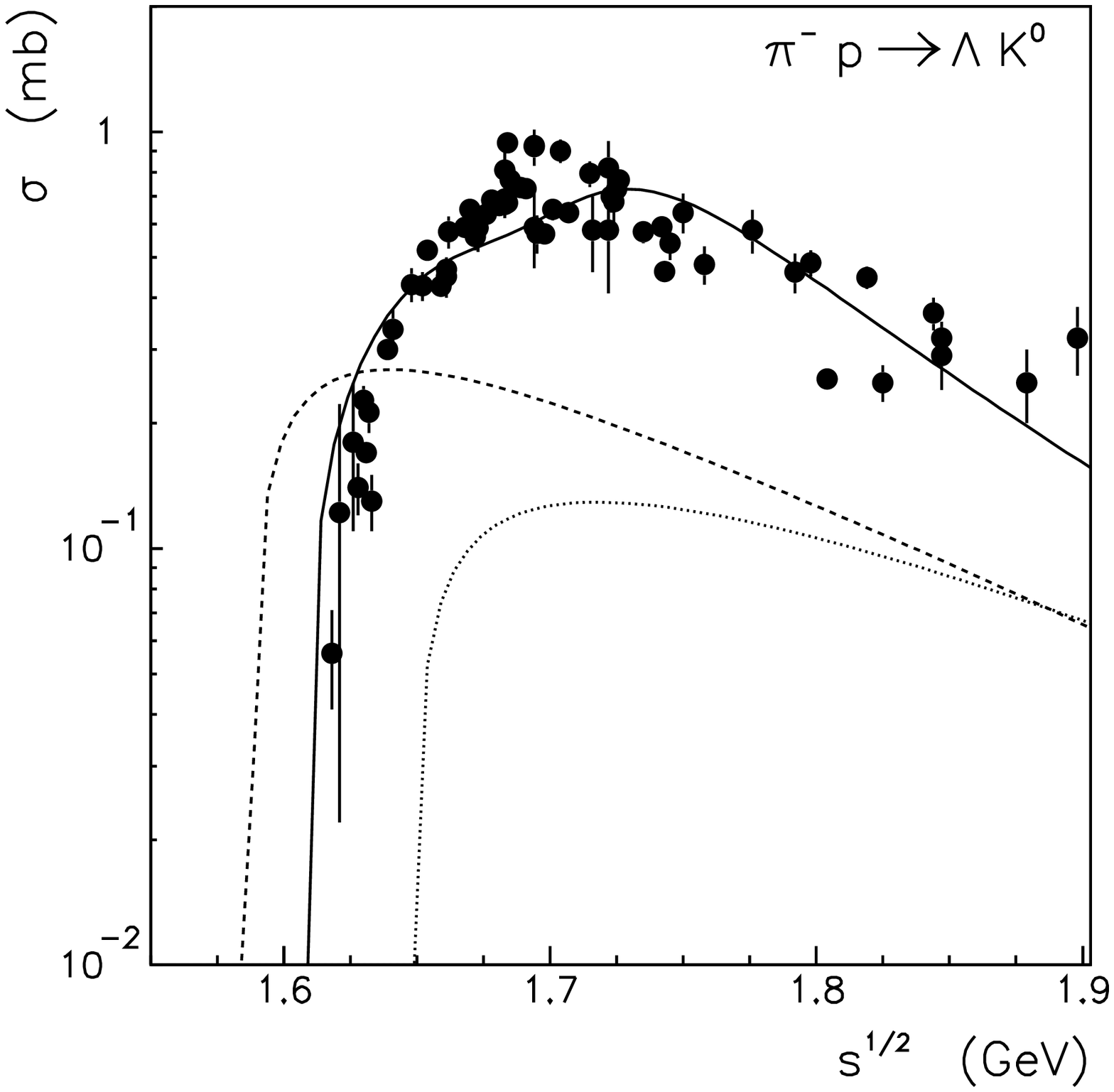,height=4.5cm,width=6cm}
\hspace{-0.5cm}
\psfig{file=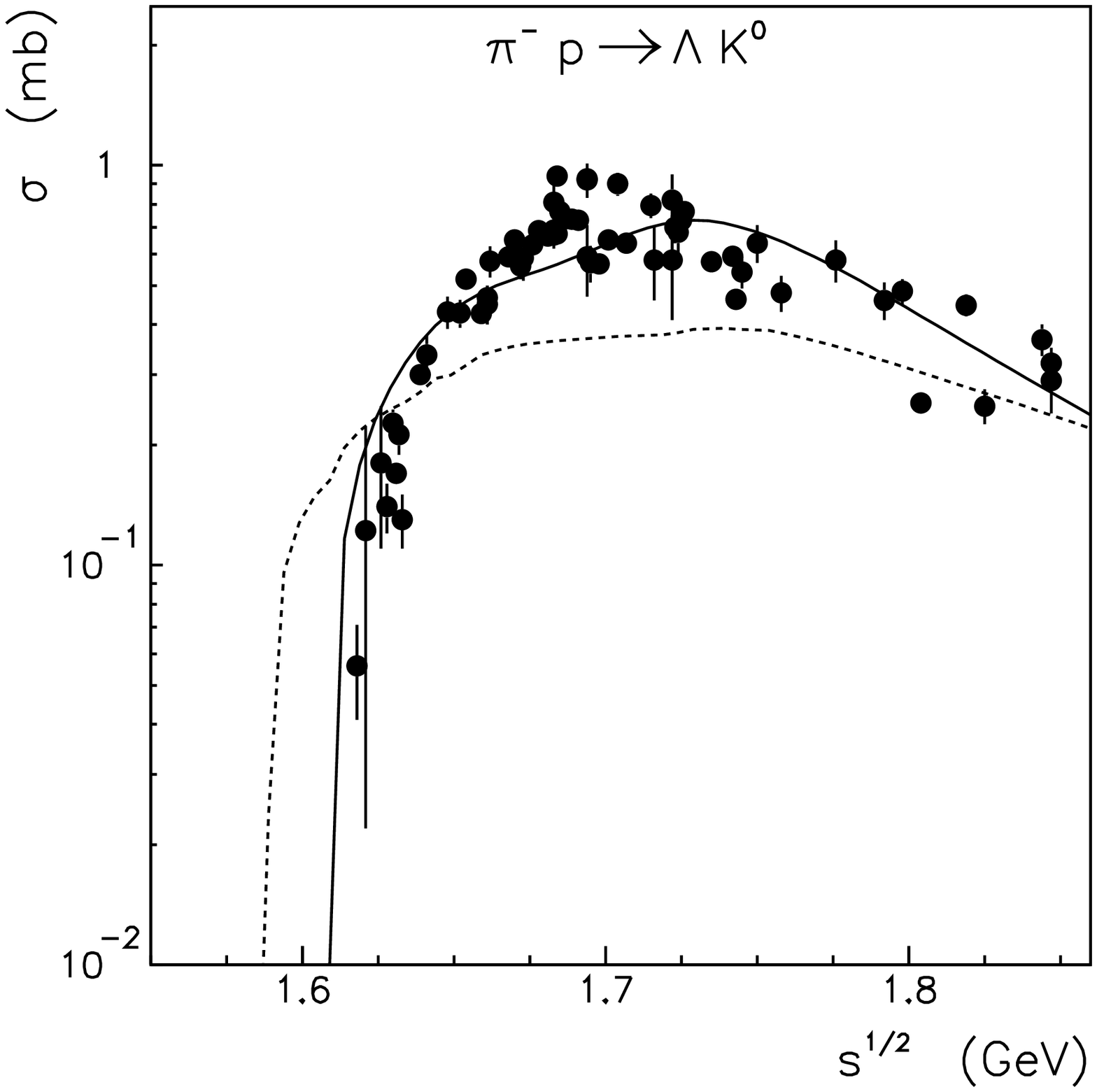,height=4.5cm,width=6cm}
\vspace{-0.2cm}
\caption{Energy dependence of the total cross sections. 
Lines in the left panel, the (solid, dashed, dotted), correspond to 
nuclear densities, $\rho_B = (0, \rho_0, 3\rho_0)$, respectively.
Lines in the right panel, the solid and dotted, correspond to 
$\rho_B = 0$, and the result averaged over the nuclear density 
distribution in Au+Au collisions at 2A GeV~\protect\cite{kprod_medium}.
The dots are data in free space~\protect\cite{kdata}, $\rho_B = 0$.
\label{k2}
}
\end{center}
\vspace{-0.5cm}
\end{figure}
%%%%%%%%%%%%%%%%%%%%%%%%%%%%%%%%%%%%%%%%%%%%%%%%%%%%

Results are shown in Fig.~\ref{k2}.  
(See caption for explanations.)
From the results
we conclude that if one accounts for
the in-medium modification of the production amplitude correctly,
it is possible to understand $K^+$ production data in heavy ion collisions
at SIS energies, even if the $K^+$-meson feels the theoretically expected,
repulsive mean field potential. The apparent failure to explain
the $K^+$ production data if one includes the purely kinematic
effects of the in-medium modification
of the $K^+$-meson and hadrons, appears to be a consequence of the
omission of these effects on the reaction amplitudes.

%%%%%%%%%%%%%%%%%%%%%%%%%%%%%%%%%%%%%%%%%%%%%%%%%%%%%%%%%%%%%%%%%%%%%%%%%%%
\section{\mb\bm{$J/\Psi$} dissociation in nuclear 
matter~\protect\cite{Tsushimaj}}
%%%%%%%%%%%%%%%%%%%%%%%%%%%%%%%%%%%%%%%%%%%%%%%%%%%%%%%%%%%%%%%%%%%%%%%%%%%
%
There is a great deal of interest in possible
signals of Quark-Gluon Plasma (QGP) formation, and $J/\Psi$ suppression
is a promising candidate as suggested by Matsui and Satz~\cite{Matsui}.
Our interest here is how much the $J/\Psi$ absorption cross sections
in hadronic dissociation processes will be modified, if the in-medium 
hadron potentials  
are included, which has never been addressed 
in QGP analyses.

\begin{figure}[thb]
\begin{center}
\vspace{-2.2cm}
\hspace*{-1cm}
\psfig{file=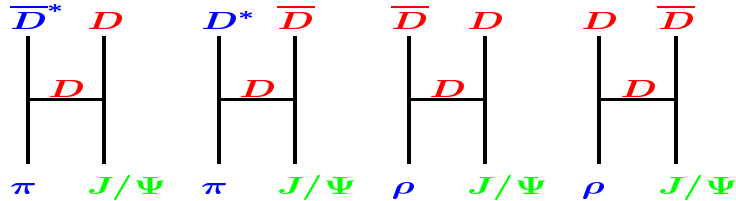,height=13cm,width=15cm}
\vspace{-8.8cm}
\caption{Processes included for $J/\Psi$ dissociation.
\label{jpsi1}
}
\end{center}
\vspace{-0.5cm}
\end{figure}

We consider the reactions involving the $J/\Psi$,
shown in Fig.~\ref{jpsi1}.
%%%%%%%%%%%%%%%%%%%%%%%%%%%%%%%%%%%%%%%%%%%%%%%%%%%%
%%%%%%%%%%%%%%%%%%%%%%%%%%%%%%%%%%%%%%%%%%%%%%%%%%
Recent calculations for the processes in free space~\cite{Mueller}, 
indicate a much lower cross sections than the necessary  
to explain the data for the $J/\Psi$ suppression.

\begin{figure}[th]
\begin{center}
\vspace{-0.2cm}
\psfig{file=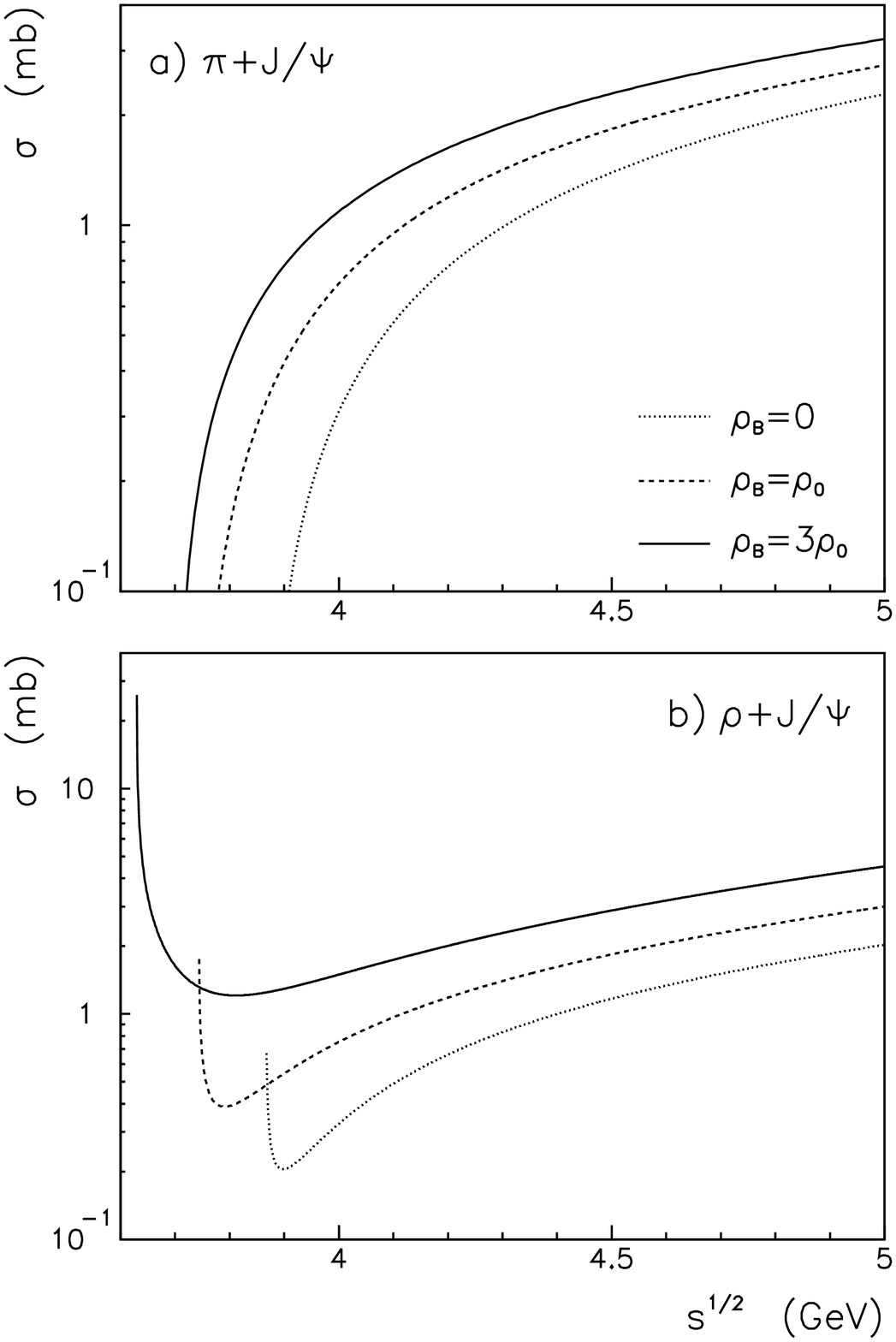,height=4.5cm,width=6cm}
\hspace{-0.5cm}
\psfig{file=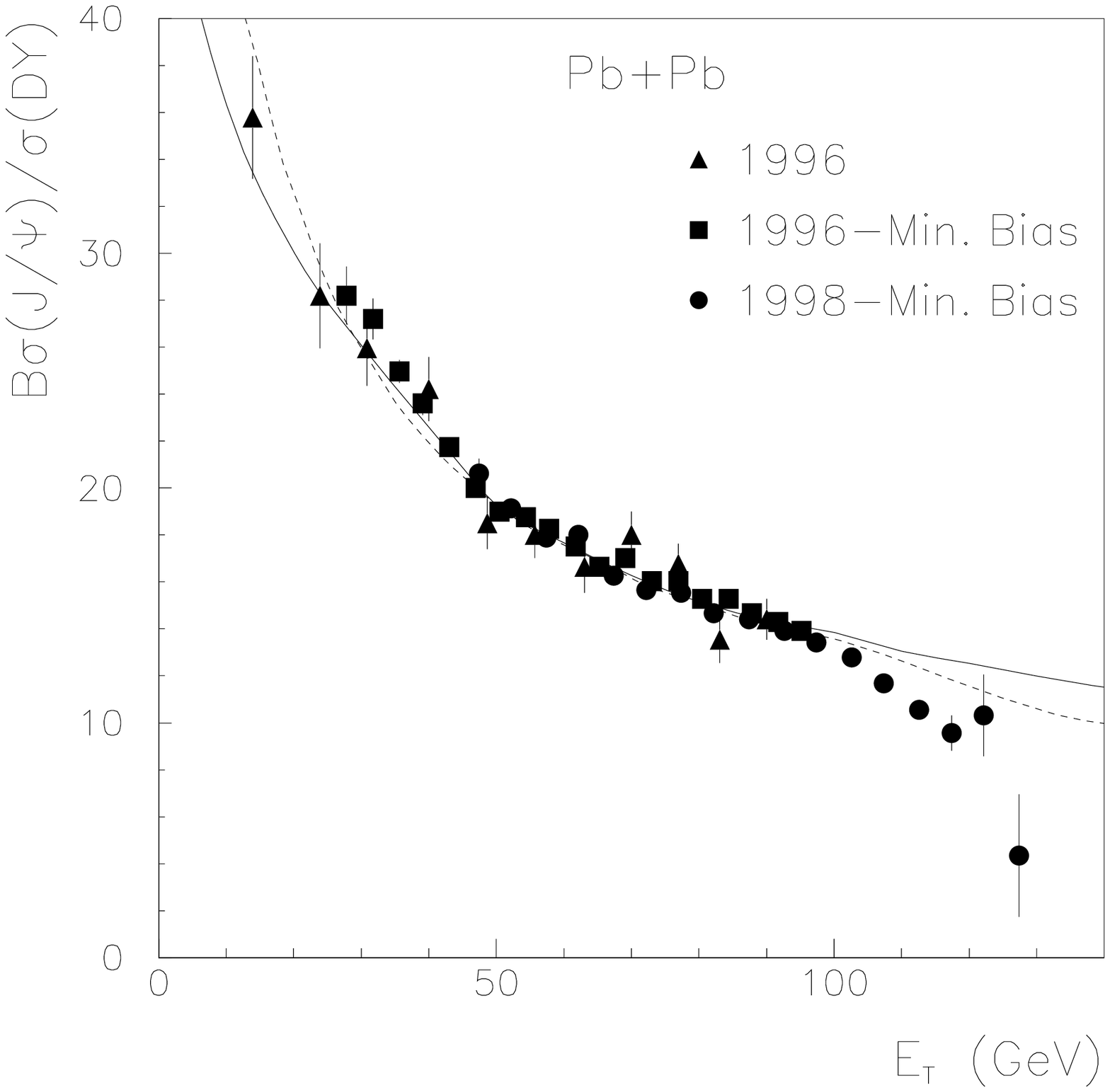,height=4.5cm,width=6cm}
\vspace{-0.2cm}
\caption{Energy dependence of the total cross section.
\label{jpsi2}
}
\end{center}
\vspace{-0.4cm}
\end{figure}

However, this situation changes when
the in-medium potentials of the charmed (and also $\rho$) 
mesons are taken into account, 
as shown in the left panel of Fig.~\ref{jpsi2}. 
%%%%%%%%%%%%%%%%%%%%%%%%%%%%%%%%%%%%%%%%%%%%%%%%%%
%%%%%%%%%%%%%%%%%%%%%%%%%%%%%%%%%%%%%%%%%%%%%%%%%%%%
Clearly, the $J/\Psi$ absorption cross sections are substantially enhanced
not only because of the downward shift of the reaction threshold, but also 
because of the in-medium effect on the reaction amplitude. 

In order to compare our results with the NA38/NA50 
data~\cite{Qm97} on $J/\Psi$ suppression in $Pb{+}Pb$
collisions, we have adopted the heavy ion model proposed in
Ref.~\cite{Capella} with the $E_T$ model from 
Ref.~\cite{Capella1}. 
%%%
%We  introduce the absorption
%cross section on comovers as function of the density of comovers,
%while the nuclear absorption cross section is taken 
%as 4.5~mb~\cite{Capella1}. 
%%%
Our result is shown in the right panel of 
Fig.~\ref{jpsi2}   
by the solid line, using the density dependent, thermally averaged
cross section~\cite{Tsushimaj}.  
The dashed line shows the result reported in Ref.~\cite{Capella1} 
using the phenomenological constant cross section, 4.5~mb. 
Both lines clearly reproduce the data~\cite{Qm97} quite well, 
including most recent results from NA50 on the ratio
of $J/\Psi$ over Drell-Yan cross sections, as a function
of the transverse energy $E_T$ up to about 100~GeV. 

It is important to note that if one neglected the in-medium modification of 
the $J/\Psi$ absorption cross section the large cross section, 
the large value for the $J/\Psi$ dissociation cross section of 4.5~mb used 
in Ref.~\cite{Capella1}, could not be justified by 
microscopic theoretical calculations, and thus
the NA50 data~\cite{Qm97} could not be described.

%%%%%%%%%%%%%%%%%%%%%%%%%%%%%%%%%%%%%%%%%%%%%%%%%%%%%%%%%%%%%%%%%%%%%%%%%%%
\section{\mb\bm{$p p \to p p \omega$ {\rm and} $p p \to p p \phi$} 
reactions at near thresholds~\protect\cite{kn,kt}} 
\begin{figure}[th]
\begin{center}
\vspace{-0.2cm}
%\hspace*{-2.5cm}
\psfig{file=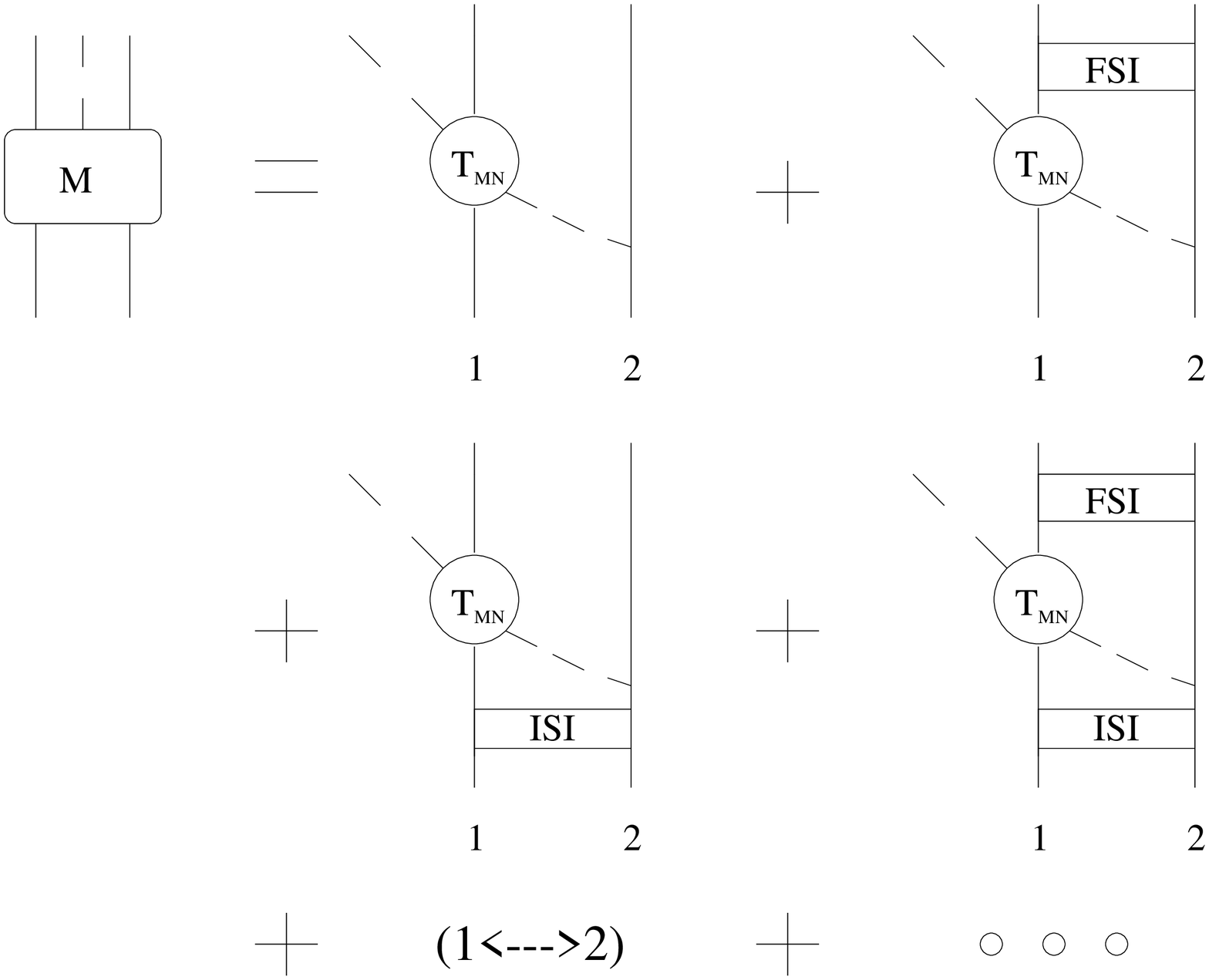,height=4cm,width=11cm}
\vspace{-0.2cm}
\caption{Amplitudes, $M$, for the $p p \to p p v\,\,(v=\omega,\phi)$ 
reactions considered in the study. $T_{MN}$, ISI and FSI stand for, 
meson-nucleon (MN) T-matrix, initial state interaction and 
final state interaction, respectively. 
\label{ppv1}
}
\end{center}
\vspace{-0.4cm}
\end{figure}

%%%%%%%%%%%%%%%%%%%%%%%%%%%%%%%%%%%%%%%%%%%%%%%%%%%%%%%%%%%%%%%%%%%%%%%%%%%
%
We report here results for the vector meson 
production reactions at near threshold, 
$p p \to p p v\,\,(v = \omega, \phi)$~\cite{kn,kt}.
Description of the model, and parameters determined by other reactions 
are given in Ref.~\cite{knvmeson}.
The vector meson production amplitude, $M$, is depicted 
in Fig.~\ref{ppv1}. 

Our main results for these vector meson production are:
\begin{enumerate}
\item{Tensor to vector coupling ratio, 
$\kappa_v \equiv f_{vNN}/g_{vNN}$, 
for the $\omega$ ($\kappa_\omega$), where a typical NN interaction 
model~\protect\cite{Bonn} sets $\kappa_\omega = 0$.}
\item{$\phi NN$ coupling constant, $g_{\phi NN}$, in connection 
with $s \sbar$ component in the nucleon wave function, 
and the OZI rule violation.}
\end{enumerate}

%%%%%%%%%%%%%%%%%%%%%%%%%%%%%%%%%%%%%%%%%%%%%%%%%%%%
%%%%%%%%%%%%%%%%%%%%%%%%%%%%%%%%%%%%%%%%%%%%%%%%%%

\begin{figure}[th]
%\vspace{-0.2cm}
\begin{center}
\psfig{file=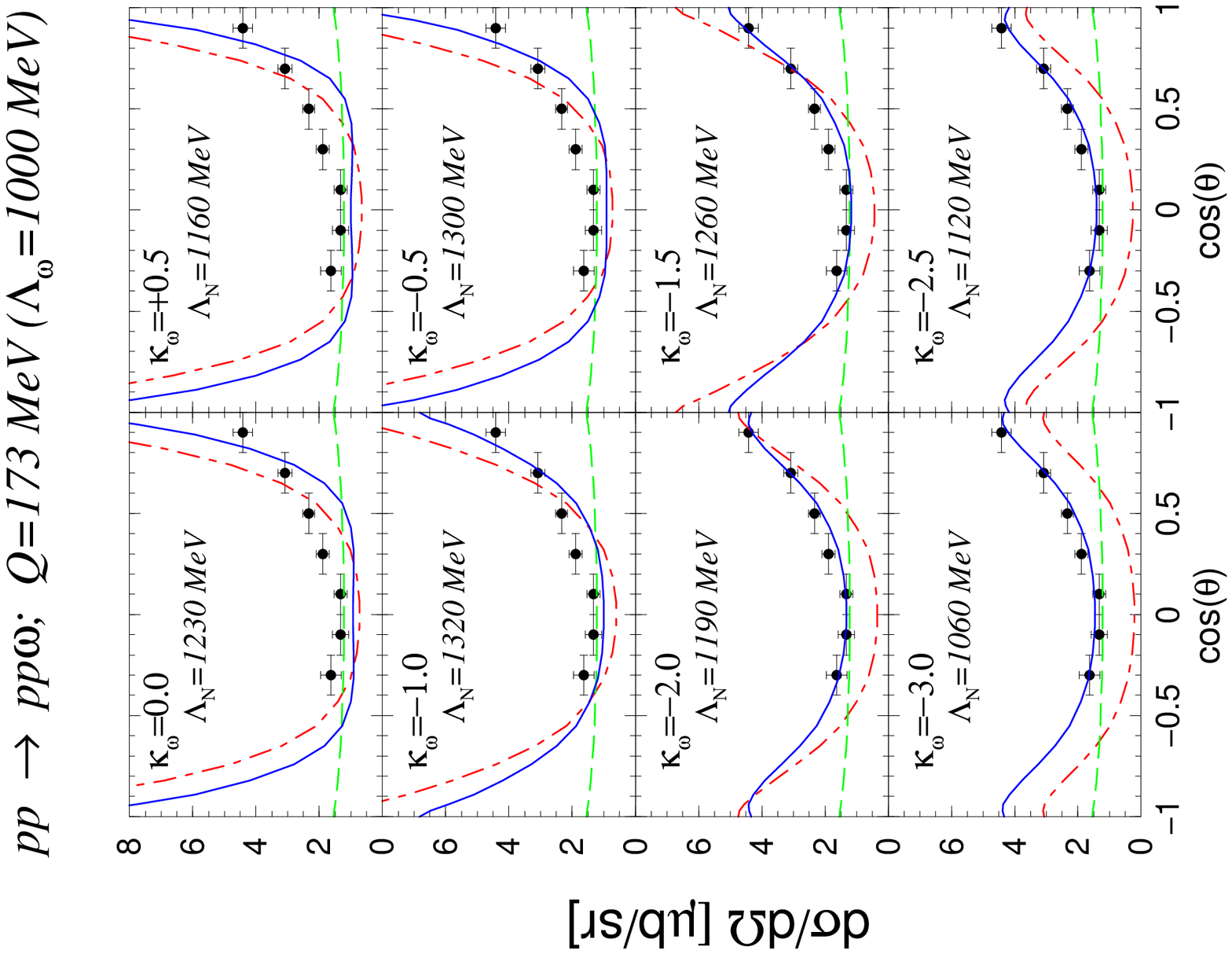,height=6.5cm,width=5cm,angle=-90}
\hspace{-1cm}
\psfig{file=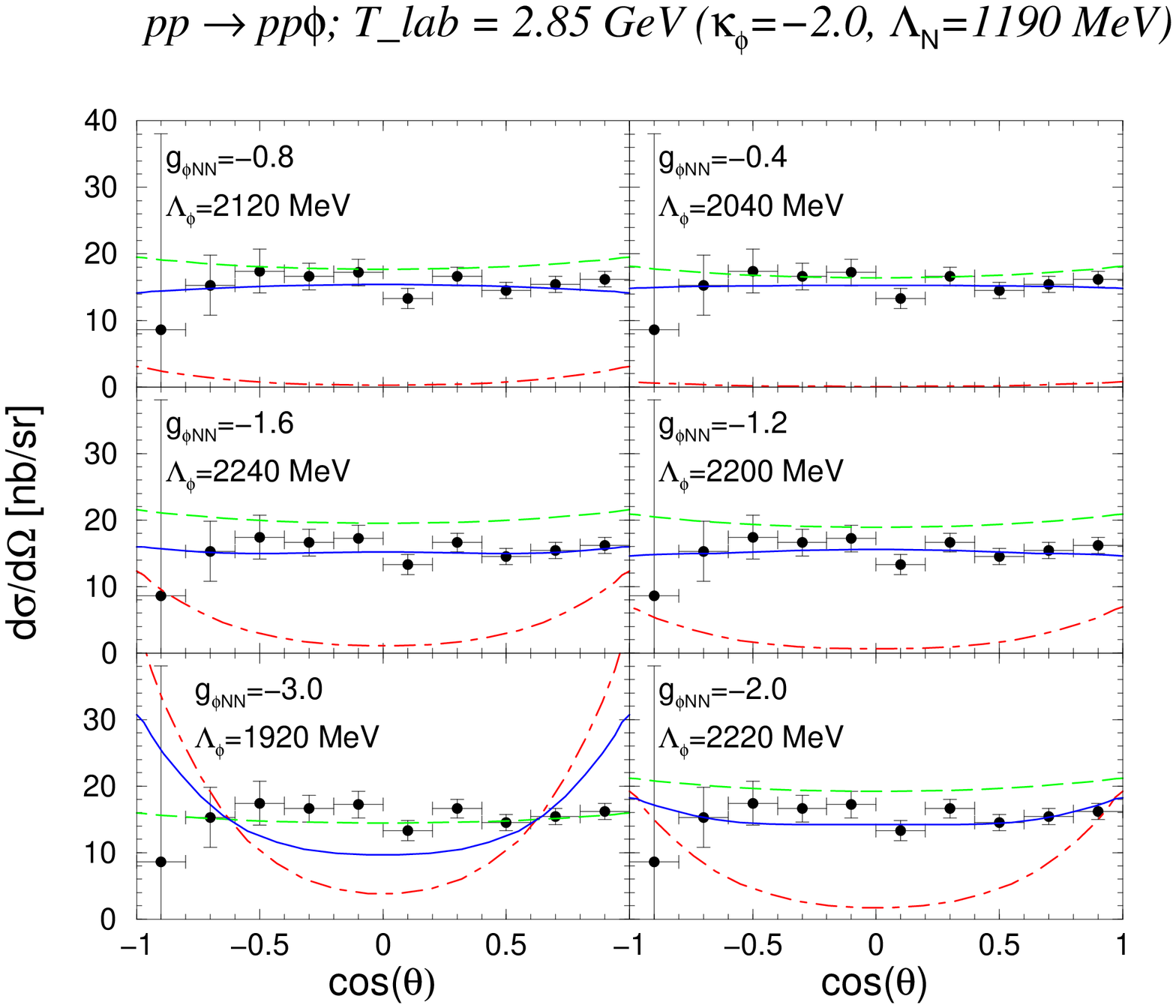,height=5.5cm,width=5cm,angle=-90}
\vspace{-0.2cm}
\caption{Angular distributions for $p p \to p p \omega$ at excess energy 
$Q = 173 MeV$ (the left panel), and for $p p \to p p \phi$ at excess 
energy $Q = 83 MeV$ (the right panel). 
They are normalized to the total cross sections, 
$\sigma(p p \to p p \omega) = 30.8~\mu b$~\protect\cite{COSY-TOF} and 
$\sigma(p p \to p p \phi) = 190~nb$~\protect\cite{DISTO}, respectively.
The (solid, dash-dotted, dotted) lines correspond to the 
(total, nucleonic, mesonic) contributions, respectively.
\label{ppv2}
}
\end{center}
\vspace{-0.6cm}
\end{figure}

Results for angular distributions  
are shown in Fig.~\ref{ppv2}, together with the experimental data 
for the $\omega$~\cite{COSY-TOF} and $\phi$~\cite{DISTO}.
Total contribution comes from following two sources, 
(1) nucleonic: the vector meson is produced from the 
$vNN$ (meson-nucleon-nucleon) 
vertex, and (2) mesonic: the vector meson is produced 
from the $v\rho\pi$ (mesonic 
vertex), where the $\pi$ and $\rho$ are attached to the two different protons.
The $\omega$ meson production is dominated equally by the 
nucleonic and mesonic, while for the $\phi$,   
the mesonic is strongly dominant. Thus, production mechanisms for these 
two vector mesons are different.
%%%%%%%%%%%%%%%%%%%%%%%%%%%%%%%%%%%%%%%%%%%%%%%%%%
%%%%%%%%%%%%%%%%%%%%%%%%%%%%%%%%%%%%%%%%%%%%%%%%%%%%
For the $\omega$ meson a large value 
$\kappa_\omega = -2.0$  
with $g_{\omega NN}^2 = (-9.0)^2 = 81.0$ fits 
best the data~\cite{COSY-TOF}, and turned out to be 
insensitive to the value of
$g_{\omega NN}$~\cite{kn,kt}, where
in Bonn NN potential model~\cite{Bonn} $\kappa_\omega = 0$  
with $g_{\omega NN}^2 \simeq 24 (4\pi) \simeq 301.6$ is used.

As for the $\phi$ meson, several parameter sets can possibly reproduce
the data~\cite{DISTO}.
Surprisingly, for $\kappa_\phi = -2.0$, a large absolute value 
$|g_{\phi NN}| = |-2.0|$, can also reproduce the data. 
This implies a large violation of the OZI rule.
However, this turned out to give a distinguishable, 
different energy dependence 
for the total cross section at near threshold~\cite{kt}.
Thus, we can distinguish if 
energy dependence of the total cross section  
is measured at near threshold.

%%%%%%%%%%%%%%%%%%%%%%%%%%%%%%%%%%%%%%%%%%%%%%%%%%%%%%%%%%%%%%%%%%%%%%%%
\vspace{-0.3cm}
\section{Conclusion}
In conclusion, there are many motivations and rich phenomena for 
performing experiments to study the changes in 
hadron properties in nuclear medium, or to study the partial 
restoration of chiral symmetry in nuclear medium.
%%%%%%%%%%%%%%%%%%%%%%%%%%%%%%%%%%%%%%%%%%%%%%%%%%%%%%%%%%%%%%%%%%%%%%%%
%
\vspace{-0.4cm}
\section*{Acknowledgments}
\vspace{-0.2cm}
The author would like to thank, D.H. Lu, K. Nakayama, K. Saito,   
A. Sibirtsev and A.W. Thomas for exciting collaborations, 
and a warm hospitality at CSSM during 
the workshop. Special thanks go to Prof. A.W. Thomas 
for many supports to attend the workshop possible.
This work was partially supported by the Fotschungszentrum-J\"{u}lich, 
contract No. 41445282 (COSY-058).

%%%%%%%%%%%%%%%%%%%%%%%%%%%%%%%%%%%%%%%%%%%%%%%%%%%%%%%%%%%%%%%%%%%%%%%%%%

%
%%%%%%%%%%%%%%%%%%%%%%
\end{document}